# Data-Driven Modelling for Harmonic Current Emission in Low-Voltage Grid Using MCReSANet with Interpretability Analysis

J. Yao, *Student Member, IEEE,* H. Yu, *Student Member, IEEE*, P. Judge, *Member, IEEE* S. Z. Djokic, *Senior Member, IEEE*

***Abstract*—Due to the diverse loads connected in grids, it is hard to insight into the complex nonlinear relationship between harmonic voltages and currents. This paper presents a novel data-driven model， Multi-Compression Refined Self-Attention Net (MCReSANet)， to construct the mapping of the highly nonlinear relationship between fundamental/harmonic voltages and currents in grids, and implements the interpretability analysis to establish the detailed correlations across various voltage and current harmonics. The MCReSANet-based model improved the Mean Absolute Error (MAE) of prediction by 10%-20% compared to the Convolutional Neural Network (CNN) and Multi-Layer Perceptron (MLP) for Finnish and German datasets, with the lowest model uncertainty. This is a crucial prerequisite for more precise interpretability analysis, by combining SHAP value-based feature importance and admittance matrix modelling. The obtained SHAP and admittance matrix shows the positive and zero sequence harmonics exhibited higher importance (SHAP value) for the Finnish and German datasets, respectively, which conforms to the characteristics of the more unbalanced three phases in Germany. Besides, it confirmed that the harmonic current prediction is more sensitive to harmonic voltages than to loads. This paper proposed a novel method to enhance the potential for predicting harmonic emissions and identifying the harmonic sources in distribution systems.**

## I. Introduction

THE widespread use of PE loads has brought significant benefits in terms of flexibility, energy conversion, and dynamic response in the power system. However, it has also led to power quality issues, particularly concerning harmonic current emissions due to the nonlinear characteristics of PE loads. Excessive harmonic current emissions can result in overloading, increased energy consumption, and equipment malfunction. Therefore, it is crucial to implement precise methods for measuring and predicting harmonic current emission by nonlinear loads so that potentially problematic harmonic currents can be detected and mitigated promptly.

The common method for obtaining the harmonic current in grids is by power quality analyzer, which includes the current measurement and harmonic detection algorithm. The common harmonic detection methods used in power quality analyzers include the Fourier transform (FT)-based method, the instantaneous reactive power theory-based method (IRPT), the wavelet transform (WT)-based method, and the neural network-based methods [1] However, the FT-based method necessitates a 200ms data window, as recommended in [2], making it less suitable for rapid response applications. The IRPT-based method can accurately detect harmonics but may struggle with triplen harmonics. WT-based methods require longer computation times. Neural network-based methods heavily rely on extensive data and data quality. Although numerous research papers have been dedicated to enhancing harmonic detection strategies [3-8], their high cost and large size remain the primary disadvantages of the currently used analysis methods due to the high requirement for the measurement hardware and its accuracy [9].

An alternative method is based on the relationship between harmonic voltage and current, as discussed in [10-13], which delves into the study of harmonic current emissions from various PE equipment, shedding light on the relationship between harmonic voltage and harmonic current emissions. The articles explore both linear and non-linear regression relationships between voltage and current harmonics, as discussed in [14]. Accordingly, several forecasting methods have been proposed based on the relationship between voltage and current harmonics, as evident in [15, 16]. However, there is a need for further improvement in prediction accuracy.

Accordingly, the traditional methods for obtaining harmonic current emissions still heavily rely on the power quality analyzer and analytical models of harmonic voltage and current. However, the high cost and weight significantly increase the inconvenience of data acquisition. Besides, it is often challenging to precisely determine the number and combination of various PE loads connected in the real grids, which adds complexity to building the analytical model of harmonic current emissions from diverse loads in the power system.

With the acquisition of a growing amount of data and advancements in intelligent artificial techniques, data-driven methods utilizing deep learning have been developed for predicting harmonic current emissions. They can capture the features and offer rapid inference capabilities from the complex environment in real grids. Besides, these methods obviate the need for traditional current measurement and harmonics

Jieyu Yao, Hao Yu, Paul Judge and Sasa Z. Djokic are with the School of Engineering, University of Edinburgh, Edinburgh, UK (e-mails: h.yu@ed.ac.uk).

detection techniques, such as FFT, WT, and IRPT-based methods. As a result, they simplify the hardware and procedures involved in measuring harmonic current emissions. Data-driven based methods were applied for predicting harmonic current emission in [17-20]. While their experimental setup only considered some specific loads (e.g., LED, CFL, Laptop, TV, etc.), which cannot fully represent the complexity of grids. Besides, further validation of the deep learning model is required using multiple datasets.

This paper successfully established the more accurate mapping of the highly nonlinear relationship between fundamental/harmonic voltages and currents in grids by using the proposed MCReSANet-based model. It also demonstrates the applicability of interpretability analysis to establish detailed correlations across various voltage and current fundamentals/harmonics. The following are the main contributions of this paper:

- The paper proposed a novel MCReSANet for building a more accurate mapping of nonlinear relationships between fundamental/harmonic voltages and currents.
- The model interpretability analysis presents the detailed relationship across various harmonic voltages and currents, by combining SHAP value-based feature importance analysis and admittance matrix modelling.
- The effectiveness of proposed method is validated by two datasets with different network characteristics.

The paper is structured as follows: Section I provides a general introduction and presents a brief literature review; Section II introduces the methodology, including the architecture of the proposed MCReSANet, loss function and benchmark; Section III introduces the training setup; Section IV demonstrates the matrix calculation for interpretability analysis, including SHAP value and admittance matrix calculation; Section V illustrates and compares the results obtained by CNN, MLP and MCReSANet, as well as conducts the interpretability analysis; Section VI presents the conclusion and outlines future work.

## II. METHODOLOGY

In this section, a comprehensive overview of the proposed MCReSANet architecture is presented. Additionally, this section includes the loss function used in the model, and the benchmarks for assessing the performance of the MCReSANet, MLP and CNN.

### A. *Architecture of the proposed Multi-Compression Refined Self-Attention Net*

In this paper, an MCReSANet-based model is proposed, as illustrated in Fig. 1. The task of establishing the nonlinear mapping between harmonic voltages and harmonic current emissions for the power system could be formulated as a regression problem, with the mathematical expression provided as follows:

$$F^{*}(\boldsymbol{\theta}) = \arg\min_{\boldsymbol{\theta}} \frac{1}{K} \sum_{i=1}^{K} \left\| F(\mathbf{x}_i \mid \boldsymbol{\theta}) - \mathbf{y}_i \right\|_2^2 \quad (1)$$

where $F^{*}(\boldsymbol{\theta})$ symbolizes the MCReSANet, with $\boldsymbol{\theta}$ being parameters of the model that are updated during the training process. The expression $F(\mathbf{x}_i|\boldsymbol{\theta})$ indicates the projected harmonic current emissions, while $\mathbf{y}_i$ represents the harmonic currents measured in the power systems.

A detailed description of the network is provided below. It comprises a multi-compression module, a single-head attention module, an MLP-Mixer module, a feature gating module, and a refinement MLP module.

*1) Multi-compression module:* The harmonic voltage data undergo preliminary processing in the feature extraction block. This block is architecturally structured with a cascade of fully connected (FC) layers, configured sequentially in ascending order of neuron counts: 32, 64, and 128. These layers are followed by a Layer Normalization (LN) [22] layer and a Gaussian Error Linear Unit (GELU) [23] activation function. The processing feature vector concludes with a final 64-neuron fully connected (FC) layer. Subsequently, the resultant $1 \times 64$ feature vector is then processed through a series of blocks, each characterized by varying compression ratios. These blocks are systematically organized with neuron counts of 4, 8, 16, 32, 64, and 128. Each block comprises an FC layer, an LN layer, and a GELU activation function, ending with a 64-neuron FC layer. The outputs from these diverse neuron-count blocks are aggregated by an element-summation feature fusion operation to form a comprehensive feature map with $1 \times 64$ dimensions. By applying varying compression ratios to feature vectors, information across distinct perceptual dimensions from harmonic voltages could be obtained.

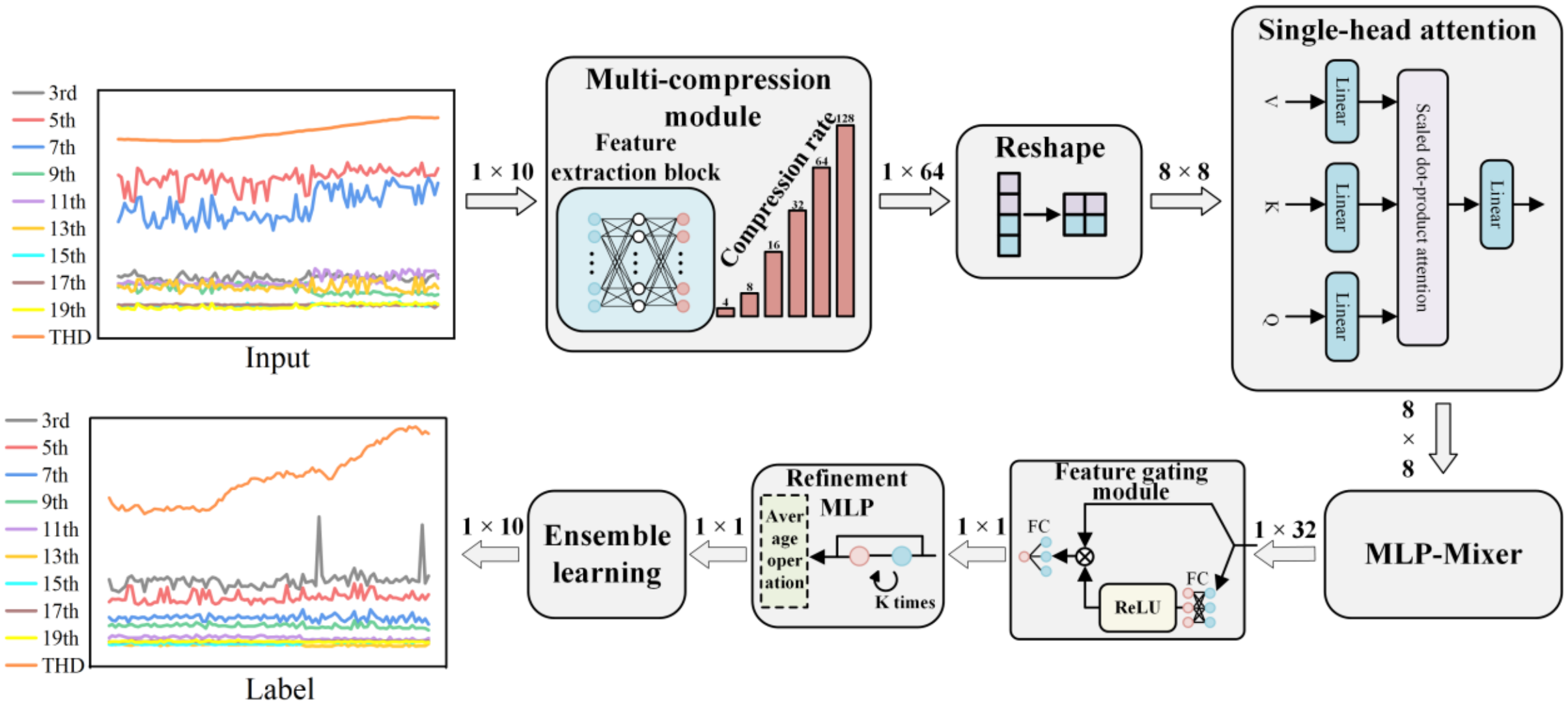

Fig. 1. Architecture of the proposed MCReSANet.

*2) Single-head attention module:* Following the aggregation, the feature is reshaped into an $8 \times 8$-dimensional matrix. This matrix is then processed through a single-head attention module [24]. The mathematical formulation for this process is outlined as follows:

$$\mathrm{Head}(\mathbf{q},\mathbf{k},\mathbf{v}) = \mathrm{Attention}\left(\mathbf{q}\mathbf{W}^{\mathbf{q}}, \mathbf{k}\mathbf{W}^{\mathbf{k}}, \mathbf{v}\mathbf{W}^{\mathbf{v}}\right)\mathbf{W}^{\mathbf{o}} \tag{2}$$

where the projections are parameter matrices $\mathbf{W}^{\mathbf{q}} \in \mathbb{R}^{d_{model} \times d_k}$, $\mathbf{W}^{\mathbf{k}} \in \mathbb{R}^{d_{model} \times d_k}$, $\mathbf{W}^{\mathbf{v}} \in \mathbb{R}^{d_{model} \times d_v}$ and $\mathbf{W}^{\mathbf{o}} \in \mathbb{R}^{hd_v \times d_{model}}$.

Reshaping a $1\times64$ feature array into an $8\times8$ format effectively segments the original feature space into smaller, distinct subspaces. Each of these subspaces is represented by a sub-feature. By applying a single attention mechanism to this structure, the network is enabled to interpret the harmonic voltage data from diverse spatial perspectives. This approach not only facilitates a multi-scale feature extraction from harmonic voltages but also reduces the dimensionality of the data, thereby decreasing computational complexity. Besides, such a methodology enhances the network's representational capacity by allowing it to capture features across various spatial scales from harmonic voltages.

*3) MLP-Mixer:* As illustrated in Fig. 2, the MLP-Mixer is a module that exclusively employs MLPs to integrate information from both the height and width dimensions within a reshaped feature vector [25]. The architecture is structured into two branches: 1) The upper branch, the height-mixing MLP, processes the columns of the input matrix $\mathbf{x}_{\mathbf{Mixer}}$. Specifically, it operates on the transposed matrix $\mathbf{x}_{\mathbf{Mixer}}^{T}$, mapping from $\mathbb{R}^H \to \mathbb{R}^H$. This MLP1 is applied uniformly across all columns. 2) The lower one, the width-mixing MLP, targets the rows of $\mathbf{x}_{\mathbf{Mixer}}$, executing a mapping from $\mathbb{R}^W$ to $\mathbb{R}^W$, and is also consistently applied across all rows using MLP1. In the proposed architecture, the output from the MLP1 block in the upper branch $\mathbf{y}_{\mathbf{upper}}$ undergoes a transposition operation. Following this, the transposed data is fused with the output from the MP1 block in the lower branch, $\mathbf{y}_{\mathbf{lower}}$, through an element-wise addition operation. This fusion process effectively integrates the features, thereby enhancing the robustness of the resultant data representation. Furthermore, the subsequent stage involves reshaping $\mathbf{y}_1$ into a single dimension and combining it with the similarly reshaped $\mathbf{y}_2$ to form $\mathbf{y}_{\mathbf{Mixer}}$, which then passes through MLP 2 to yield the output of the module. By leveraging skip connections and transposition operations, the MLP-Mixer could efficiently capture both local and global features of reshaped maps.

*4) Feature gating module*: The feature gating module acts as a gating mechanism, employing the Rectified Linear Unit (ReLU) [26] activation to weigh features in its output. The mathematical expression of the module is as follows,

$$\mathbf{y}_{\mathbf{Fg}} = \mathbf{W}\left(\mathrm{ReLU}\left(\mathbf{W}_{\mathbf{Gate}}\mathbf{x}_{\mathbf{Fg}} + \mathbf{b}_{\mathbf{Gate}}\right) \odot \mathbf{x}_{\mathbf{Fg}}\right) + \mathbf{b} \tag{3}$$

where $\mathbf{x}_{\mathbf{Fg}}$ and $\mathbf{y}_{\mathbf{Fg}}$ are inputs and outputs respectively. $\mathbf{W}_{\mathbf{Gate}}$, $\mathbf{W}$, $\mathbf{b}_{\mathbf{Gate}}$ and $\mathbf{b}$ represent the weights and biases, respectively, which are updated during the backpropagation process. Moreover, $\odot$ is the Hadamard product operator.

This module ensures a focus on essential characteristics of harmonic voltages while minimizing other factors. Furthermore, by maintaining non-negative values, the network adaptively adjusts its intermediate features, enhancing its ability to handle complex voltage data and optimizing its effectiveness in modelling tasks.

*5) Refinement MLP:* Within the refinement MLP module, data dimensions are consistently maintained at $1 \times 1$. The network undergoes iterative optimization over $K$ MLP iterations, incorporating residual connections to enhance training stability. Moreover, by averaging the aggregated outputs after each iteration, the stability of the resultant harmonic current emissions could be further improved.

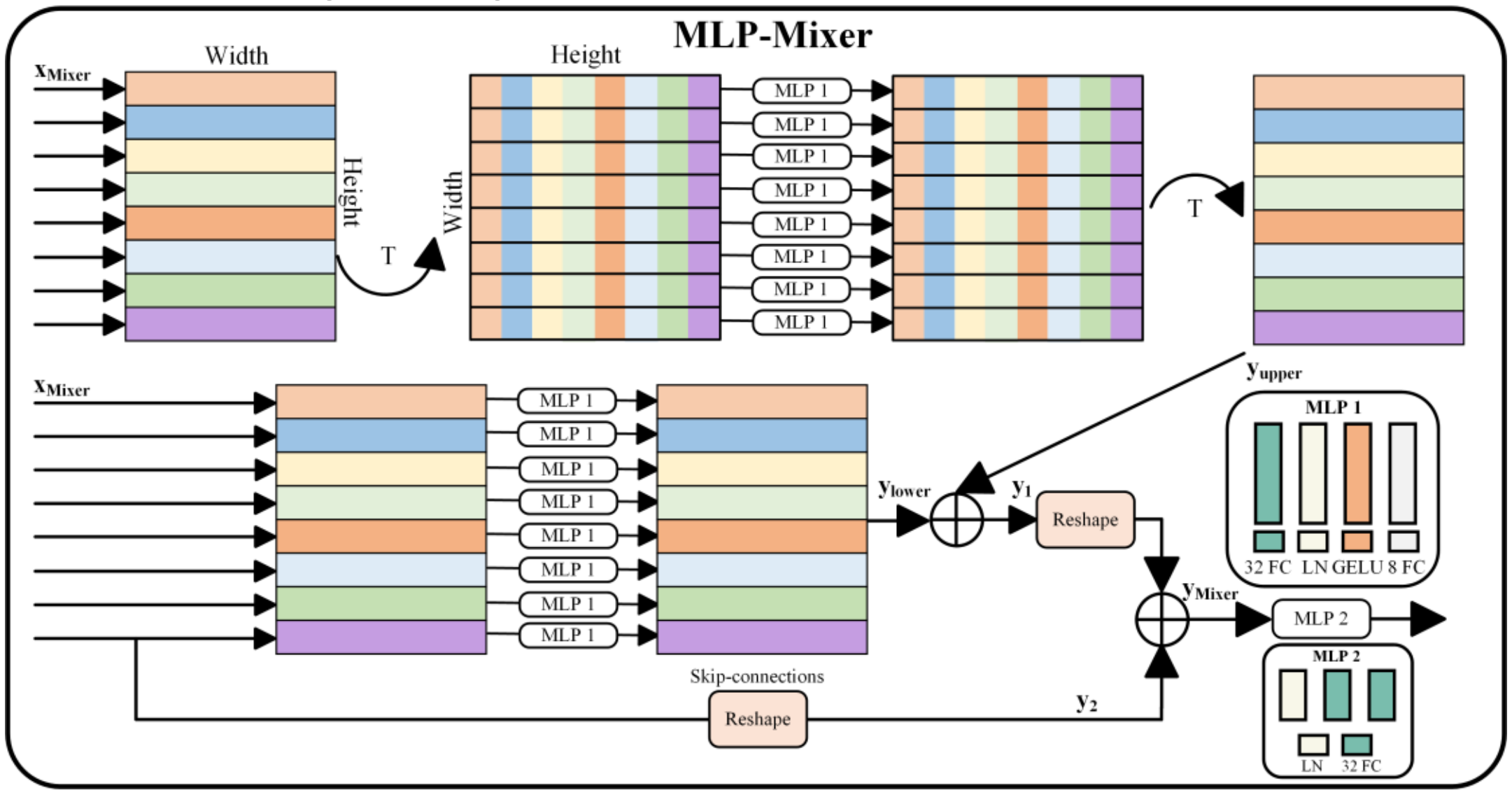

Fig. 2. The Architecture of the MLP-Mixer.

*6) Ensemble learning:* An ensemble learning approach is used to integrate harmonic currents and THD. Because harmonic current orders are primarily related to load characteristics and are largely independent of each other. By employing separate feature extraction networks for specific harmonic currents or THD, more accurate regression models between harmonic voltages and currents in the low-voltage grid could be established. Furthermore, this approach allows for a more precise capture of individual harmonic currents and THD characteristics, with reduced interference between them, thereby bolstering the robustness of the model across various load types.

## B. Loss Function

The establishment of the nonlinear mapping between harmonic voltages and harmonic current emissions can be regarded as a regression problem. Consequently, the mean squared error (MSE) loss function is used in the paper and is defined as follows:

$$\mathcal{L}_{loss} = \mathcal{L}_{MSE} + \alpha \|\boldsymbol{\theta}\|^2 \quad (4)$$

where the first term is the MSE loss and the second term is an $l_2$ regularization with regularization parameter $\alpha$.

## C. Benchmark

To assess the performance of the proposed MCReSANet, we adopted two deep learning benchmarks: multi-layer perceptron (MLP) and convolutional neural network (CNN).

*1) MLP*: MLP shares architectural similarities with the proposed MCReSANet. Structurally, the MLP initiates with an input layer of 10 neurons. This is succeeded by a hidden layer encompassing 32 neurons, which is further followed by subsequent layers containing 64 and 128 neurons respectively. After peaking at 128 neurons, the architecture undergoes a reduction process: it first retracts to 64 neurons and subsequently to 32 neurons. The architecture culminates in an output layer with 10 neurons. Within this structure, ReLU serves as the activation function inserted between each pair of successive fully connected layers. This function introduces non-linearity, helping establish a nonlinear mapping between harmonic voltages and harmonic current emissions.

*2) CNN*: The CNN network begins with a 1D convolution using a kernel size of 3, followed by max pooling and ReLU activation. Subsequent convolutional layers escalate channel counts from 16 to 32, and then 32 to 64, each accompanied by max pooling and ReLU. After the convolutional operations, the data is flattened and passed through fully connected layers that transition from 128 neurons to 64, then 32, and finally 10 output neurons. ReLU activation is used consistently between the fully connected layers.

# III. Training Setup

## A. Data Collection

In terms of the data collection in Finland, the data was measured in 2018 by a power quality analyzer at one PCC in Helsinki, Finland [27], which supplies around 3 households. The selected measurement point was in a residential area, and without local generation and electric vehicles. The average harmonic voltages and currents were recorded every 10 seconds, and the measurement lasted 8 weeks in autumn. For data collection in Germany, the power quality analyzer measured data at one PCC in 2013 in the residential area. The capacity of connected loads is 10 times higher than that in the Finnish measurement point without local generation and electric vehicles. The average harmonic voltages and currents were recorded every 1 minute, and the measurement lasted one year.

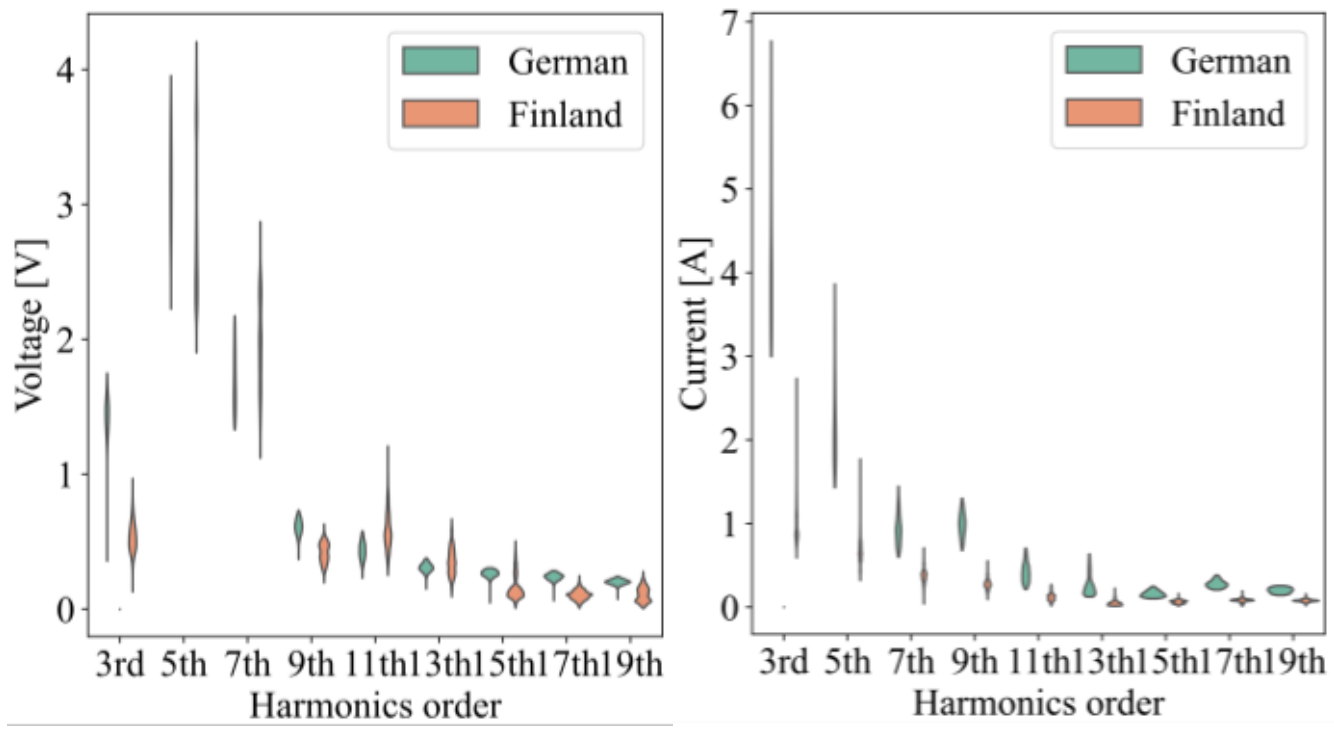

(a) Voltage (b) Current

Fig. 3. The distribution of (a) voltage and (b) current harmonics in Finland and Germany grid.

The selected Finnish and German dataset lack phase angle information, and there are several interruptions within the measurement period due to measurement reconfiguration. Due to these interruptions, those datasets are not suitable for the time series deep learning model.

## B. Data Characteristics

By analysing the datasets from Finland and Germany, the voltage and current harmonics in the German and Finnish grids exhibit distinct characteristics, as shown in Fig. 3. It is evident that the amplitudes of the 3rd and 9th voltage harmonics in the German grid are greater than in the Finnish grid, indicating more pronounced three-phase imbalance issues in Germany. Additionally, harmonic current emissions in the German grid consistently exceed those in Finland. The fluctuations and range in harmonic current emissions within the German network are considerably larger than those in the Finnish network. This phenomenon can be explained by the fact that the loads connected at the German measurement point are ten times larger than the Finnish measurement point. Therefore, there are clear distinctions in terms of harmonic distribution, magnitude, and fluctuation range between the two selected datasets from Finland and Germany.

## C. Data preprocessing

In this paper, the one-week data for the first phase in the three-phase system is sufficient for preparing the dataset for model training. As mentioned before, the Finnish dataset was measured in autumn, so one week in winter was selected for the German datasets to emphasize the difference in load types compared to the Finland datasets. Because autumn and winter usually represent the lowest and highest load demand. The following are the main steps for data preprocessing.

*1) Outlier Removal:* Since transient events, measurement noise, and data transmission errors can occur during long-term harmonic measurements, it is essential to remove outliers to enhance dataset quality. Filtering outliers is accomplished by applying the 95th percentile rule, where outliers are defined as data points exceeding the value at the percentile index.

$$P_i = 95\% \times (N+1) \tag{5}$$

$$95\,percentile: h_{outliers} \geq C_{P_i} \tag{6}$$

where $P_i$ is the percentile index, representing the 95$^{th}$ percentile position within the sorted ascending dataset, $N$ is the total number of points, $h_{outliers}$ is the outliers, $C_{P_i}$ is the data value at the 95$^{th}$ percentile. By applying (5) and (6), the outliers can be removed.

*2) Data standardization:* Standardizing data can effectively mitigate the influence of outliers and data with substantial variations. In this paper, Z-score standardization [28, 29] was utilized to maintain uniform scaling across the data, thereby bolstering the robustness and convergence attributes of the ensuing training procedure (7).

$$x_{std} = \frac{x_i - \mu}{\sigma} \tag{7}$$

where $x_i$ is the original data point; $\mu$ is the mean value of data; $\sigma$ is the standard deviation of data.

*3) Train and test sets:* The datasets from Finland and Germany were partitioned into training and test sets at about a ratio of 0.85:0.15. In this paper, the train and test sets are prepared for single input and single output (SISO) system as well as the multiple input and multiple output (MIMO) system. The SISO system assumed the prediction of specific order harmonic current would not be impacted by the other order harmonic voltages (e.g., the 3$^{rd}$ harmonic current most significantly depends on 3$^{rd}$ harmonic voltage). While the MIMO system assumed there are interactive impacts across various orders of harmonic voltage and current. Thus, for the SISO system, the input and output data should be the pair of single elements within the sets of (8) with the same order (e.g., 3$^{rd}$ harmonics voltage and 3$^{rd}$ harmonic current). For MIMO, the input and output data should be multiple pairs of multiple elements within the sets. In this paper, the MIMO system is classified as 11 inputs and outputs system (11_MIMO) including fundamental components, and 10 inputs and outputs system (10_MIMO) without fundamental components.

$$\begin{aligned} Input_i &\in \{v_1, v_3, v_5, v_7, v_9, v_{11}, v_{13}, v_{15}, v_{17}, v_{19}, thd_v\} \\ Output_j &\in \{i_1, i_3, i_5, i_7, i_9, i_{11}, i_{13}, i_{15}, i_{17}, i_{19}, thd_i\} \end{aligned} \tag{8}$$

Where $\{v_1, v_3, v_5, v_7, \ldots v_{19}\}$ and $\{i_1, i_3, i_5, i_7, \ldots i_{19}\}$ are the sets including fundamental and odd order harmonic voltage/current from 1$^{st}$ to 19$^{th}$, $thd_v$ and $thd_i$ is the THD for voltage and current;

### D. Evaluation Index

To evaluate the performance of harmonic current emission prediction methods, three indexes are used in this paper: R-squared ($R^2$), mean absolute error (MAE), and root mean squared error (RMSE), calculated as (9-11).

$$R^2 = 1 - \frac{\sum (y_i - \hat{y}_i)^2}{\sum (y_i - \bar{y})^2} \tag{9}$$

$$MAE = \frac{1}{n}\sum_{i-1}^{n} |(y_i - \hat{y}_i)| \tag{10}$$

$$RMSE = \sqrt{\frac{1}{n}\sum_{i-1}^{n} (y_i - \hat{y}_i)^2} \tag{11}$$

where $y_i$ is the actual data, $\hat{y}_i$ is the predicted data, $\bar{y}$ is the mean value of predicted value, $n$ is the number of points. $R^2$ is an index that measures the proportion of the variance in the dependent variable that is predictable from the independent variables and can evaluate the goodness-of-fit of the model, i.e., if the $R^2$ is equal to 1, it means the predicted data is the same as the actual data. *MAE* and *RMSE* can reflect the prediction accuracy, while the RMSE gives more weight to larger errors and outliers.

### E. Model Training and Implementation

Deep learning models were implemented using a laptop equipped with an NVIDIA GeForce RTX 4080 12GB GPU and 32GB RAM and powered by an Intel Core i9-13980HX processor. All models were developed using PyTorch 2.0.1 and Python 3.10. Moreover, we employed a step decay training strategy with an initial learning rate of 0.001. An early stopping mechanism was utilized, limiting training limited to a maximum of 100 epochs. Additionally, a batch size of 64 was used, and l2 regularization with a coefficient of 1×10-5 was incorporated to mitigate overfitting. The Adam optimizer was employed for optimization.

## IV. Interpretability Analysis Matrix Calculation

To further interpret the relationship across various fundamental/harmonic voltages and currents in grids, the SHAP value and admittance matrix are calculated. The SHAP value can indicate the contribution of each input on prediction which is related to the partial derivative of harmonic current concerning harmonic voltage. While the harmonic admittance matrix can build the physical significance of admittance/impedance across voltage and current harmonics. The method of combining and comparing the SHAP value and harmonic admittance matrix can insight into the detailed correlations between input (voltage information) and output (current information) in grids (e.g. the harmonic contribution from the utility or customer side can be evaluated). Besides, it should be noted, before calculating the SHAP value and admittance matrix, standardization is implemented for increasing consistency and comparability of various harmonic orders, since the magnitude of fundamentals and harmonics are usually lower for higher order.

### A. SHAP value Computation

The SHAP value is proposed based on the game theory, representing the contributions of features to predictions. The SHAP value is defined as the weighted average of marginal contributions for each feature across all possible coalitions. The formula for calculating the SHAP values for each feature is as (12-13) [30]. Accordingly, the SHAP value is obtained by comparing the total margin contribution with and without the feature. As the calculation complexity is high, the calculation method is usually estimated by Monte Carlo integration [31]. Besides, the SHAP value is considered related to the partial derivative of output (current) concerning input (voltage) as (14) (e.g., [32, 33]).

$$\varphi_i(j) = \sum_{\mathbf{S} \subseteq \mathbf{S} \cup j} \omega(|\mathbf{S}|) \cdot [v(\mathbf{S} \cup j) - v(\mathbf{S})] \tag{12}$$

$$\omega(|\mathbf{S}|) = \frac{|\mathbf{S}|!(p - |\mathbf{S}| - 1)!}{p!} \tag{13}$$

$$\varphi_i(j) = f\left(\frac{\partial Output_i}{\partial Input_j}\right) \tag{14}$$

where $j$ is a feature belongs to the feature set, which includes each order harmonic voltage; $\varphi_i(j)$ is the SHAP value of feature $j$; $|\mathbf{S}|$ is the size of subset '$\mathbf{S}$'; $p$ is the number of features; $\omega(|\mathbf{S}|)$ is the weight; $v(\mathbf{S} \cup j)$ and $v(S)$ are the marginal contribution for the set $\mathbf{S} \cup j$ and $S$, which are the difference between prediction value and exception of prediction value. $Input_j$ and $Input_i$ are one of the features $j$ and $i$ within input and output sets.

### B. *Harmonic Admittance Matrix*

The impedance modelling establishes the mathematical relationship between fundamental/harmonic voltages and currents. The equivalent circuit of harmonic impedance is drawn as in Fig.4. It should be noted that the total impedance in the grid should be classified into the impedance at the utility and customer side. Thus, the impedance of utility and customer sides can be obtained as (15-16). Since the Finnish and German dataset only has the magnitude information measured at PCC without phase angle information, the harmonic impedance modelling can only be conducted for the magnitudes of impedance at the customer side.

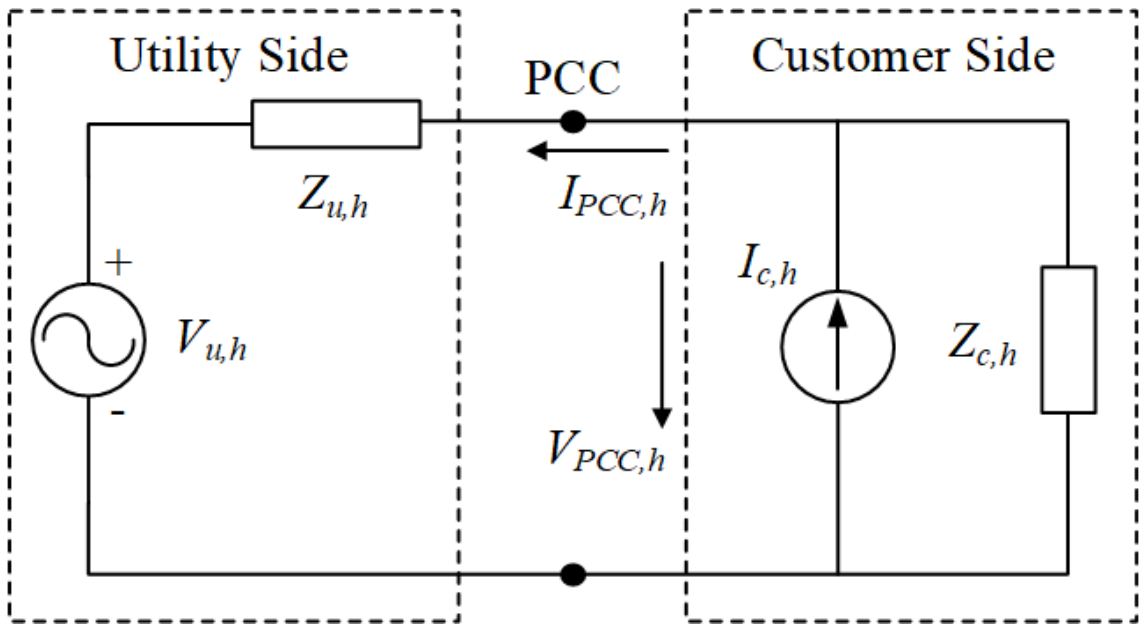

Fig. 4. The equivalent circuit of harmonic impedance at PCC.

$$Z_{c,h} = \frac{V_{PCC,h}}{I_{PCC,h}} \tag{15}$$

$$Z_{u,h} = \frac{V_{u,h} - V_{PCC,h}}{I_{PCC,h}} = \frac{V_{u,h}}{I_{PCC,h}} - Z_{c,h} \tag{16}$$

Where $Z_{c,h}$ and $I_{c,h}$ are the impedance and current source for each order harmonic at customer side; $Z_{u,h}$ and $V_{u,h}$ are the impedance and voltage source for each order harmonic at utility side. $V_{PCC,h}$ and $I_{PCC,h}$ are the voltage and current measured at PCC for each order harmonic.

Since the model training setup is implemented for the SISO and MIMO systems, those two systems should be also considered in for admittance matrix calculation for keeping the consistency.

*1) SISO and MIMO system:* For the SISO system, there is only self-admittance between the single input and output. The admittance of the SISO system is as (17). In terms of the MIMO system, it assumed there are self-admittance and mutual admittance, so the admittance matrix calculation is as (18).

$$|i_h| = |Y_h| \cdot |v_h| \tag{17}$$

$$\begin{bmatrix} |i_1| \\ |i_3| \\ \vdots \\ |i_{19}| \\ |thd_i| \end{bmatrix} = \begin{bmatrix} |Y_{11}| & |Y_{13}| & \cdots & |Y_{1,19}| & |Y_{1,thd_v}| \\ |Y_{31}| & |Y_{33}| & \cdots & |Y_{3,19}| & |Y_{3,thd_v}| \\ \vdots & \vdots & \vdots & \vdots & \vdots \\ |Y_{19,1}| & |Y_{19,3}| & \cdots & |Y_{19,19}| & |Y_{19,thd_v}| \\ |Y_{thd_i,1}| & |Y_{thd_i,3}| & \cdots & |Y_{thd_i,19}| & |Y_{thd_i,thd_v}| \end{bmatrix} \begin{bmatrix} |v_1| \\ |v_3| \\ \vdots \\ |v_{19}| \\ |thd_v| \end{bmatrix} \tag{18}$$

Where $|i_h|$ and $|v_h|$ represent the single input and output within voltage and current sets; $|Y_h|$ represents the admittance magnitude for SISO system; $[|v_1|, |v_3|, \dots |v_{19}|, |thd_v|]$ and $[|i_1|, |i_3|, \dots |i_{19}|, |thd_i|]$ are multiple inputs and outputs within the voltage and current sets; $[|Y_{ij}|]$ are the admittance magnitude matrix including self admittance and mutual admittance for MIMO systems.

*2) least squares (LS) method for admittance matrix calculation:* Since the voltage and current matrix $\mathbf{V_j}$ and $\mathbf{I_i}$ are singular, it is hard to calculate the admittance matrix $\mathbf{Y_{ij}}$ directly. In this paper, the optimized estimation of the admittance matrix can be obtained by utilising LS method as (19).

$$\min f(\mathbf{Y_{ij}}) = \|\mathbf{Y_{ij}} \mathbf{V_j} - \mathbf{I_i}\|^2 \tag{19}$$

Where $\mathbf{Y_{ij}}$ is the admittance magnitude matrix; $\mathbf{V_j}$ and $\mathbf{I_i}$ are the input (voltage) and output (current) matrix.

## V. Results

### A. *Harmonic Current Emission Prediction*

The accuracy (MAE and RMSE) for SISO, 11_MIMO and 10_MIMO systems by using MCReSA, CNN and MLP are shown in Fig. 5. The results indicate that the harmonic current information cannot be predicted accurately for the SISO system, while the errors of all methods for 11_MIMO and 10_MIMO systems are within acceptable limits for the German and Finnish datasets. Besides, all three methods and systems are not able to predict the fundamental component accurately, but the addition of fundamental components into model inputs increased the prediction accuracy. This indicates that the fundamental current is not dominantly impacted by fundamental and harmonic voltages. Instead, it is impacted by connected load much more significantly. Therefore, it is capable of establishing mappings for highly nonlinear relationships between harmonic voltage and current except for fundamental components.

Notably, the results from MCReSANet show significantly higher accuracy, not only in total prediction but also in predicting each order of harmonic current and THD, both for the 11_MIMO and 10_MIMO systems. In terms of the 11_MIMO system, the total MAE of MCReSANet improved by 15% and 19% compared to CNN and MLP for the Finland dataset, and by 9% and 10% compared to CNN and MLP for

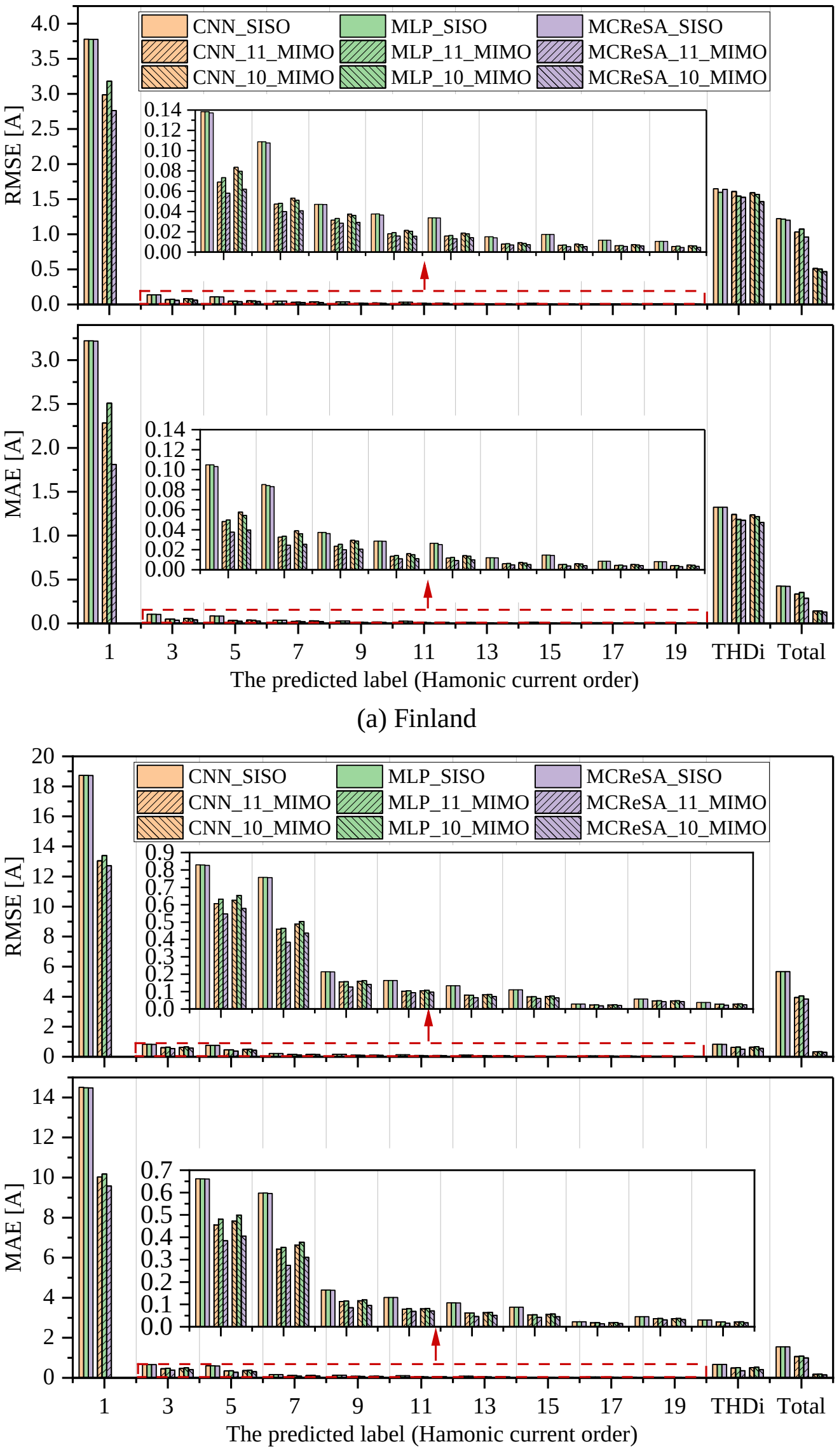

Fig. 5.Comparison of MAE and RMSE obtained by MLP and MCReSANet for each order harmonic current emission in Finland and German network.

TABLE II

THE IMPROVEMENT OF MCRESA COMPARED WITH CNN AND MLP IN TERMS OF RMSE AND MAE

| Country | Methods | Improvement of MCReSA [%] | | | | | |
|---|---|---|---|---|---|---|---|
| | | SISO | | 11_MIMO | | 10_MIMO | |
| | | CNN | MLP | CNN | MLP | CNN | MLP |
| Finland | RMSE [A] | 1.74 | 1.28 | 7.00 | 10.69 | 8.77 | 7.14 |
| | MAE [A] | 0.64 | 0.34 | 14.71 | 18.91 | 8.45 | 7.80 |
| Germany | RMSE [A] | 0.10 | 0.11 | 6.61 | 8.17 | 10.68 | 13.17 |
| | MAE [A] | 0.13 | 0.02 | 8.80 | 10.37 | 16.24 | 19.46 |

the Germany dataset. Also, the total RMSE of MCReSANet improved by 7% and 11%, as well as 7% and 8% compared to CNN and MLP for Finland and Germany datasets respectively, as in Table II. This affirms that the proposed MCReSANet excels in extracting harmonic features for diverse and complex grid-connected loads in the low-voltage network, which is effective in improving accuracy, even in two datasets with different characteristics, as analysed in Section III.

Furthermore, the uncertainty of CNN, MLP, and MCReSANet for SISO, 11_MIMO and 10_MIMO systems is compared in Fig. 6, by repeating training in five times. The results confirm that all models exhibit stable performance for two datasets, especially in the case of the German network, but MCReSANet exhibits exceptional stability compared to CNN and MLP for all harmonics. The enhanced stability in uncertainty analysis can be attributed to the implementation of the Refinement MLP module, which utilizes loop iteration and averaging operations to achieve methodical convergence and precision in the results. However, the 10_MIMO exhibited a higher uncertainty than that of 11_MIMO, indicating that the addition of fundamental current could decrease the model uncertainty, which means the fundamental voltage also contributes to the current prediction.

Accordingly, MCReSANet can predict most accurately when there are 11 inputs and outputs. In Fig. 7, the scatter density chart of the 11_MIMO system compares the actual and predicted values of each order of harmonic current using the proposed MCReSANet for Finland and Germany datasets. The $R^2$ values fall within the range of 0.71 to 0.91 for Finland and 0.74 to 0.89 for Germany. These values indicate the high fitness of the MCReSANet model and its high accuracy in predicting each order harmonic current. However, due to the higher magnitudes and fluctuations of harmonic current emissions in the German dataset, as discussed in Section III, the $R^2$ values for harmonic current prediction in Germany are slightly lower than those for Finland. It is worth noting that the $R^2$ for the 7$^{th}$ harmonic current is the lowest, while the $R^2$ for the 15$^{th}$ harmonic current is the highest, for both the Finland and Germany datasets. This suggests that the source of the 7th harmonic current in the datasets could be the greatest diversity, making it more challenging for MCReSANet to capture 7th harmonic information. Fig. 8 shows the time series results by MCReSANet for two datasets, which indicates that the predicted results are close to the original data, but some spikes in datasets cannot be predicted. Since spikes may be caused by unexpected system behavior or external interference (e.g., harmonic currents are impacted by the change of connected load instead of harmonic voltages), the regression relationship for spikes data is hard to be trained.

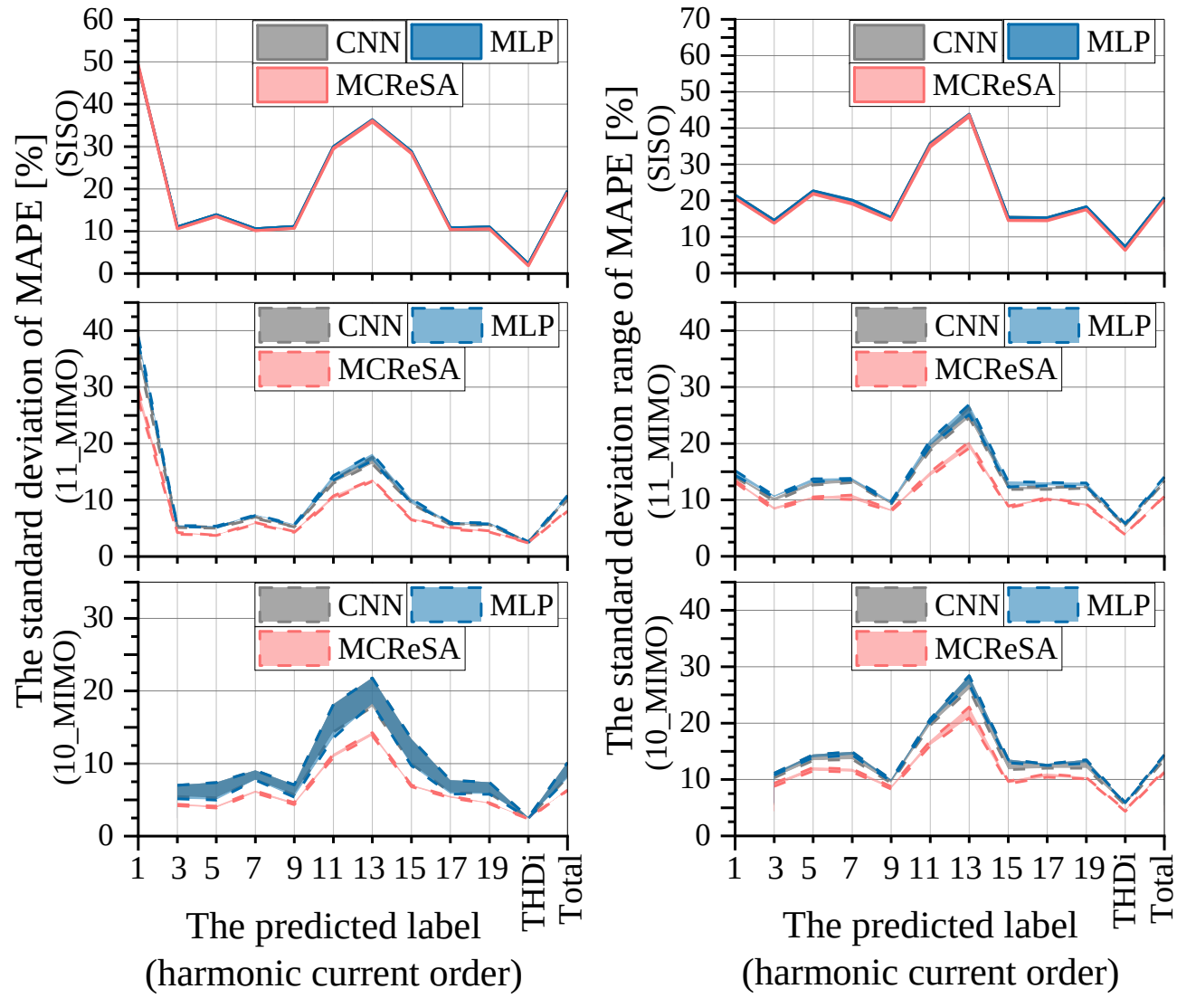

(a) Finland (b) Germany

Fig. 6. Uncertainty analysis of CNN, MLP and MCReSA, in case of training model in repeated five times

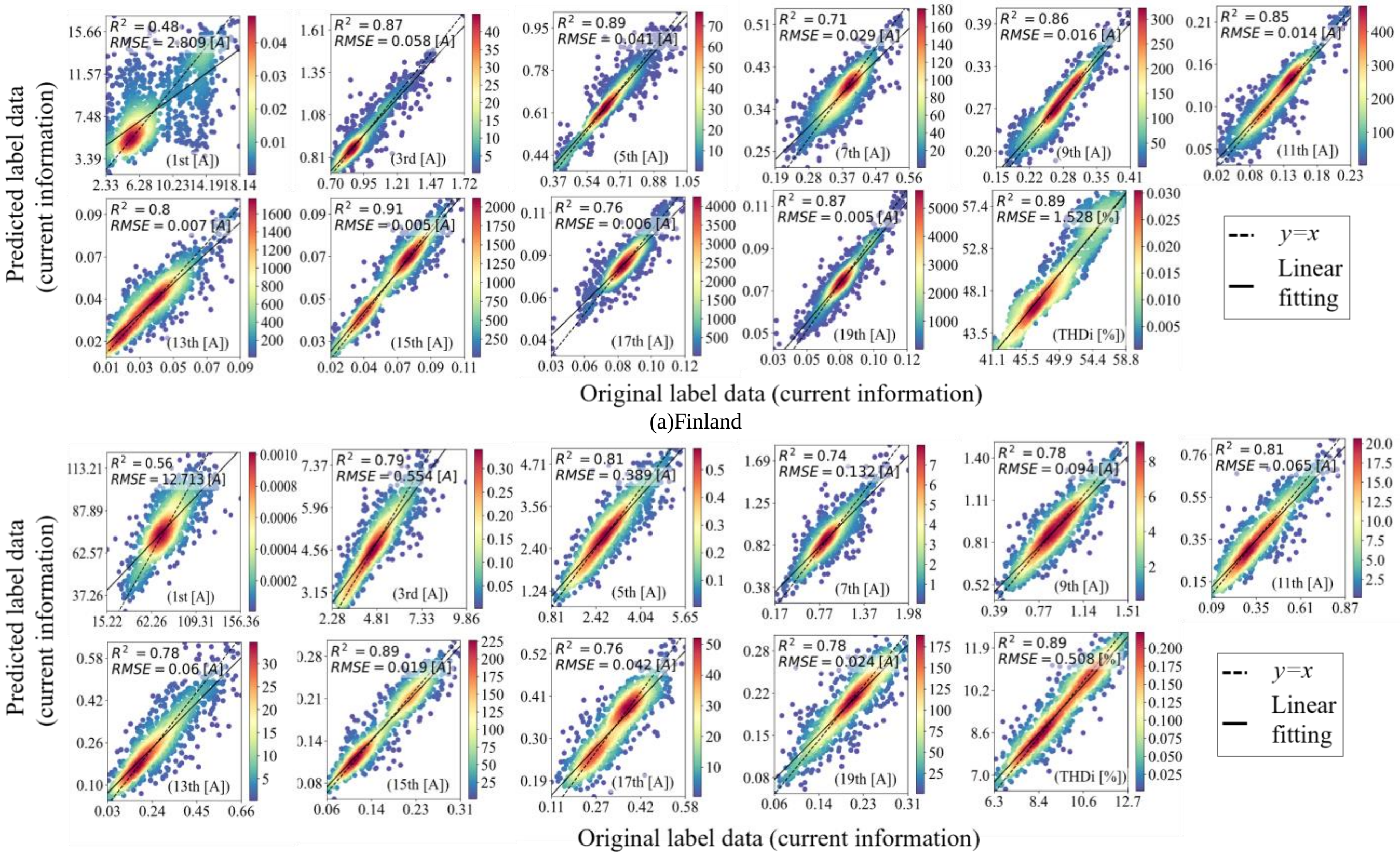

(b) Germany

Fig. 7. Scatter density plot for comparing actual and predicted harmonic current by using MCReSANet for Finland and Germany datasets.

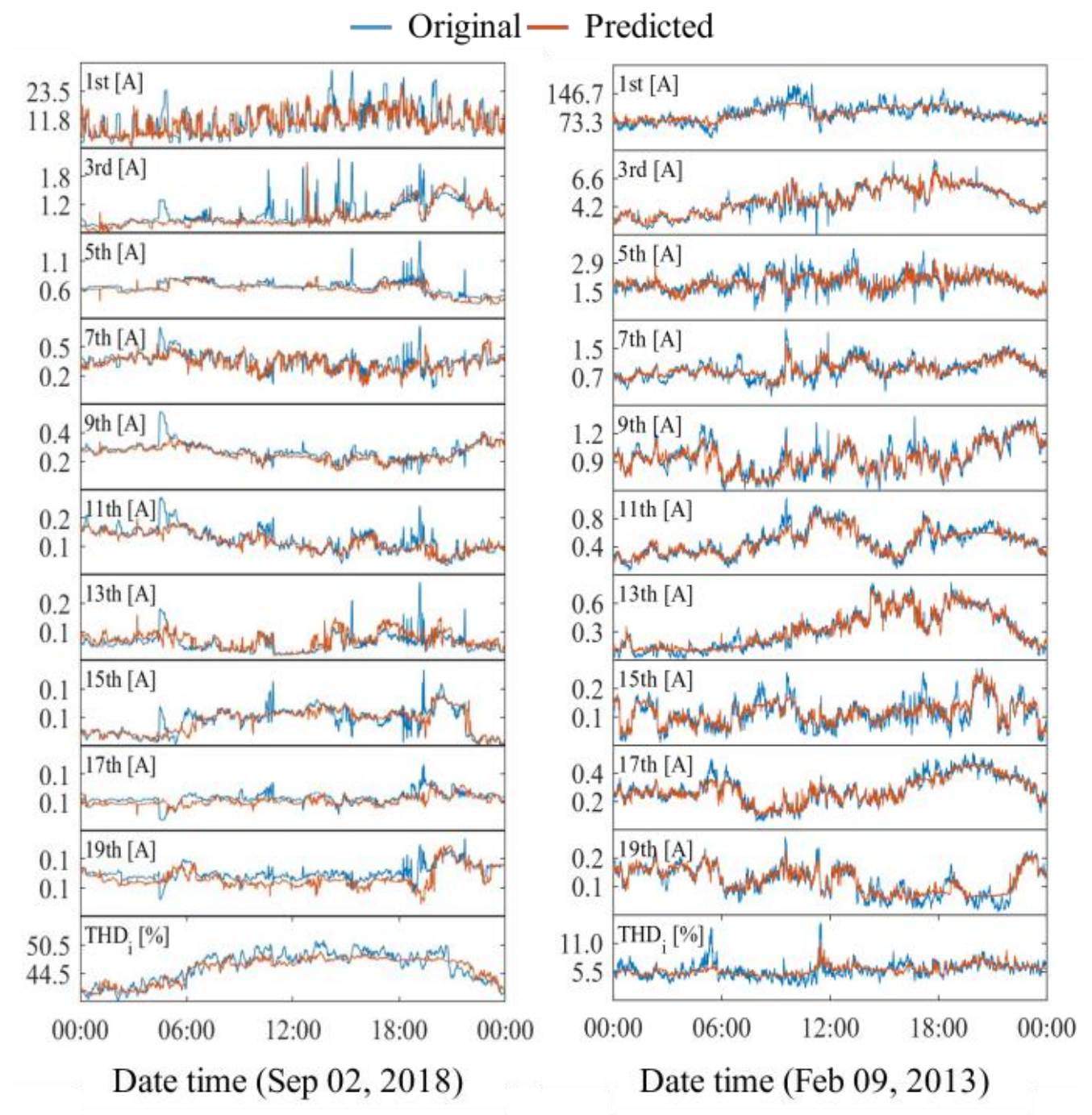

(a) Finland (b) Germany

Fig. 8. Time series results for Finland and Germany datasets.

## *B. SHAP Value-Based Interpretability Analysis*

The feature importance, analysed by computing SHAP values for each input feature, can indicate the sensitivity of detailed correlation between harmonic voltages and currents. The higher SHAP value means the higher feature importance and sensitivity. Since the SISO system cannot obtain accurate predictions, it is not necessary to implement interpretability analysis for the SISO system. Fig. 9 presents the SHAP values of each harmonic voltage and current for 10_MIMO and 11_MIMO systems, as well as the standardized admittance matrix for 11_MIMO system (the inputs and outputs are standardized before admittance matrix calculation as mentioned in Section IV). The admittance matrix of the 10_MIMO system is included in that of the 11_MIMO system, since the admittance matrix for 10_MIMO is very close to that for 11_MIMO. The results indicate that the SHAP values for the Germany dataset are much higher than those for the Finland dataset, which suggests that the output harmonic information is more sensitive to the input voltage features in the German dataset. This heightened sensitivity can be attributed to the more distinct characteristics of higher magnitude and fluctuation of harmonic currents in the German dataset.

In terms of the Finland dataset, it is evident that the positive sequence harmonics (i.e., $7^{th}$, $13^{th}$, and $19^{th}$) exhibit considerably higher importance in influencing harmonic current prediction compared to other orders. Furthermore, the $5^{th}$, $7^{th}$, $9^{th}$, $13^{th}$, $19^{th}$ and THD reveal higher SHAP values for the diagonal order of harmonic voltages and currents. While others exhibited the cross-correlation between various harmonic orders (e.g., $7^{th}$ harmonic voltage also has a high importance on the $9^{th}$ harmonic current prediction). For the Germany dataset, these positive sequence harmonic voltages do not exhibit high importance. Instead, the positive sequence harmonic currents are primarily impacted by the 3rd, 9th and 15th harmonic voltages. Furthermore, similar to the Finland dataset, the 3rd, 7th, 9th, 13th, 17th and 19th reveal higher SHAP values for the diagonal order of harmonic voltages and currents. While there are also interactive impacts across various harmonic orders.

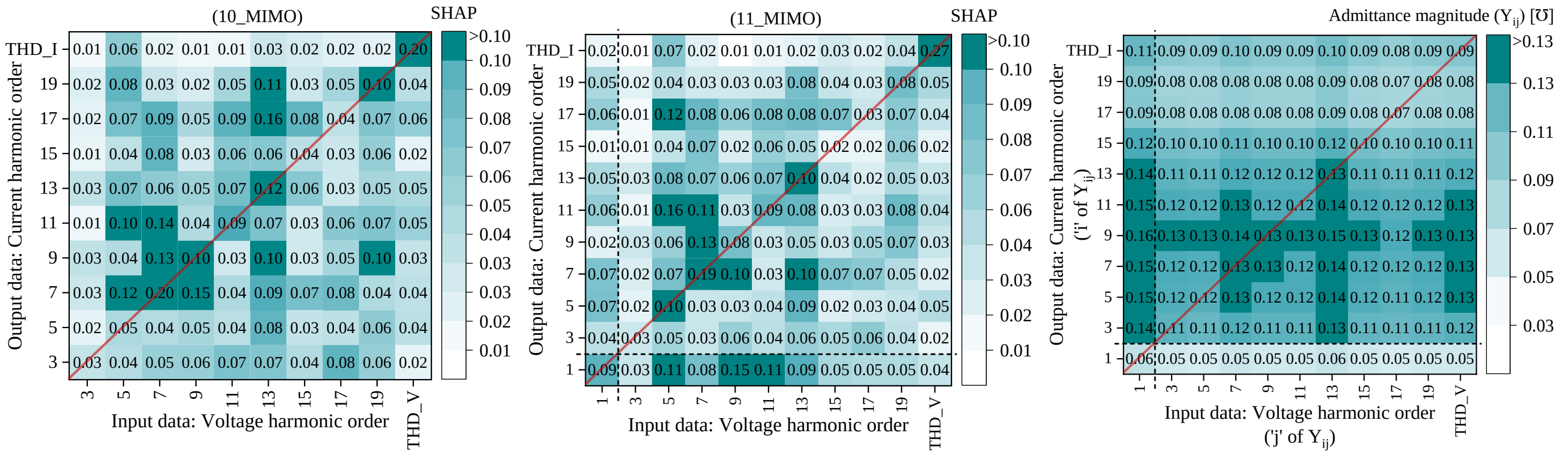

(a) Finland

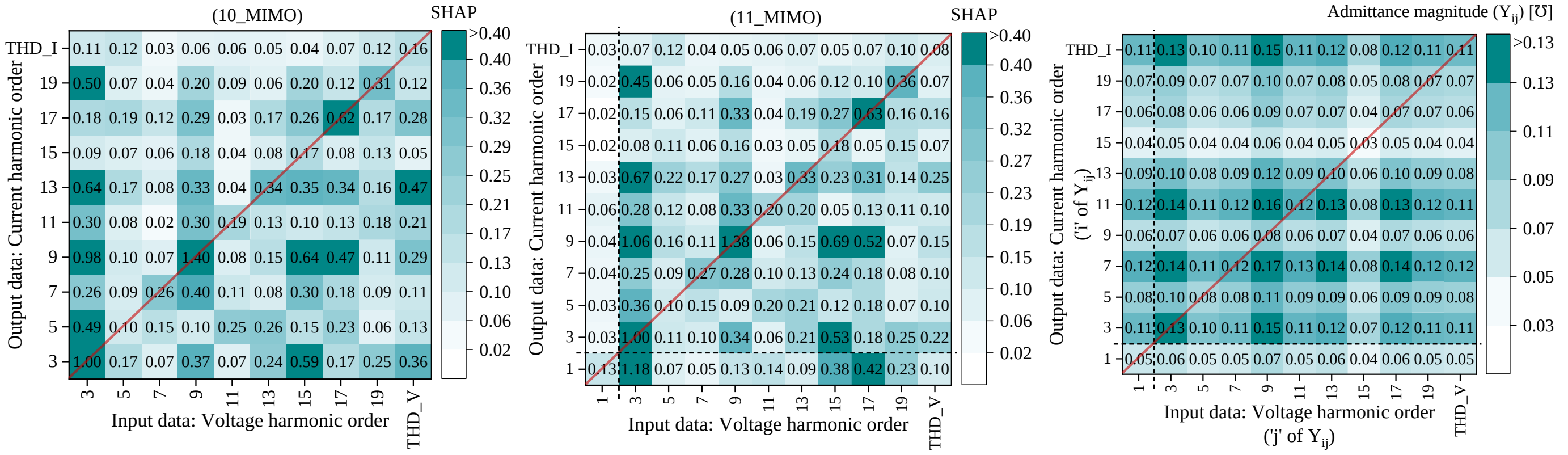

(b) Germany

Fig. 9. SHAP values and standardized admittance matrix for each harmonic voltages and currents.

Accordingly, it becomes evident that harmonics in electrical systems can exhibit complex and nonlinear interactions among themselves. Positive and zero sequence harmonics have a dominant impact in the German and Finnish datasets respectively. Thus, the Finnish measurement point exhibits a balanced grid with well-behaved loads since the phase sequence of positive sequence harmonics is the same as that of the fundamental component [34]. This also indicates that the German measurement point exhibits a more severe three-phase unbalance than the Finnish measurement point. This aligns with the Germany and Finland network characteristics in 2013 and 2018 respectively (the period of datasets), i.e., there is a larger number of connected loads at the German measurement point, and there is more harmonic propagation from the respective region, which has been affected by a greater amount of DEG and nonlinear loads [35, 36].

Moreover, it found that after adding fundamental voltage to inputs, the SHAP value of most harmonic voltages and currents decreased for both Finland and Germany datasets, which confirms the fundamental voltage contributes to the ouput current prediction. However, the results also show that the SHAP value of fundamental voltage on current prediction is very low, especially for the Germany dataset, which is the lowest among other orders, which indicates that the harmonic current prediction is more sensitive to harmonic voltages. Therefore, the fundamental current cannot be predicted only by fundamental/harmonic voltages since the fundamental components much more significantly depend on the change of connected loads. While the fundamental voltage contributes to the current prediction but not that important as harmonic voltages.

In terms of the standardized admittance matrix, it can compare the impedance across various orders of harmonic voltage and current. In terms of Finnish admittance matrix, the admittance related to fundamental current are the lowest, but there are just slight differences among the admittance for other harmonics. For German admittance matrix, the admittance related to fundamental current and 15th harmonic currents are the lowest. This indicates that the loads/impedances related to the fundamental current are the highest for two datasets. This also aligns with the above results that the fundamental current prediction more highly depends on the change of connected load than harmonic voltages. Therefore, the combination of SHAP values and admittance matrix enables us to analyse the detailed regression relationships between each order harmonic voltage and current more comprehensively.

## VI. Conclusion

This paper has shed light on establishing the more accurate mapping of the highly nonlinear relationship between harmonic voltage and current emission in low voltage networks by a data-driven method, even if the grid-connected loads in the network are highly complex and diverse. The proposed MCReSANet significantly improves the accuracy of constructing the model by 15% and 19% compared to the CNN and by 9% and 10% compared to the MLP in terms of MAE for two selected Finnish and German datasets. This serves as the foundation for accurately analysing the detailed relationship for each order of harmonic voltage and current by a model interpretability analysis method – the combination of the SHAP value-based feature importance analysis and admittance matrix modelling. The SHAP and admittance matrix indicate there are interactive impacts among all harmonics, but the positive sequence harmonics and zero sequence harmonics are dominant for selected Finland and Germany datasets respectively, which conforms to their various network characteristics, i.e., the larger amount of load demand connected and a more complex network context (e.g., nonlinear loads and DEG) for the German measurement point than Finnish. Besides, it confirms that the harmonic current prediction is much more sensitive to harmonic voltages than to loads. Therefore, this paper overcomes the difficulties in building analytical models and establishes correlations between the voltage and current harmonics, which is beneficial to more effective management for optimizing power quality in diverse grid environments. Future work will involve implementing the prediction and interpretability analysis by the time series deep learning model to insight into the time-related correlations between voltage and current harmonics.